\documentclass[prb,twocolumn,showpacs,superscriptaddress]{revtex4}
\usepackage{graphicx,dcolumn,amsmath,bm}

\begin{document}

\title { 
Calculated spin-orbit splitting of all diamond-like and zinc-blende semiconductors: 
Effects of $p_{1/2}$ local orbitals and chemical trends
}

\author{\firstname{Pierre} \surname{Carrier}} 
\author{\firstname{Su-Huai} \surname{Wei}}
\affiliation{National Renewable Energy Laboratory, Golden CO 80401, USA}

\date{\today}

\begin{abstract}
We have calculated the spin-orbit (SO) splitting $\Delta_{SO}=\epsilon
(\Gamma_{8v}) - \epsilon (\Gamma_{7v})$ for all diamond-like group IV and
zinc-blende group III-V, II-VI, and I-VII semiconductors using the full
potential linearized augmented plane wave method within the local
density approximation. The SO coupling is included using the second
variation procedure, including the $p_{1/2}$ local orbitals. The
calculated SO splittings are in very good agreement with available
experimental data. The corrections due to the inclusion of the
$p_{1/2}$ local orbital are negligible for lighter atoms, but can be as large as
$\sim$250 meV for 6$p$ anions. We find that (i) the SO splittings
increase monotonically when anion atomic number increases; (ii) the SO
splittings increase with the cation atomic number when the compound is
more covalent such as in most III-V compounds; (iii)
the SO splittings decrease with the cation atomic number when the
compound is more ionic, such as in II-VI and the III-nitride compounds;
(iv) the common-anion rule, which states that the variation of
$\Delta_{SO}$ is small for common-anion systems, is usually obeyed, especially for
ionic systems, but can break down if the compounds contain second-row elements
such as BSb;
(v) for IB-VII compounds, the 
$\Delta_{SO}$ is small and in many cases negative and it does not follow the rules 
discussed above.
These trends are explained in terms
of atomic SO splitting, volume deformation-induced charge renormalization,
and cation-anion $p$--$d$ couplings.
\end{abstract}


\maketitle
\section{Introduction}
 \begin{figure}[htb]
   \includegraphics*[width=7.5cm]{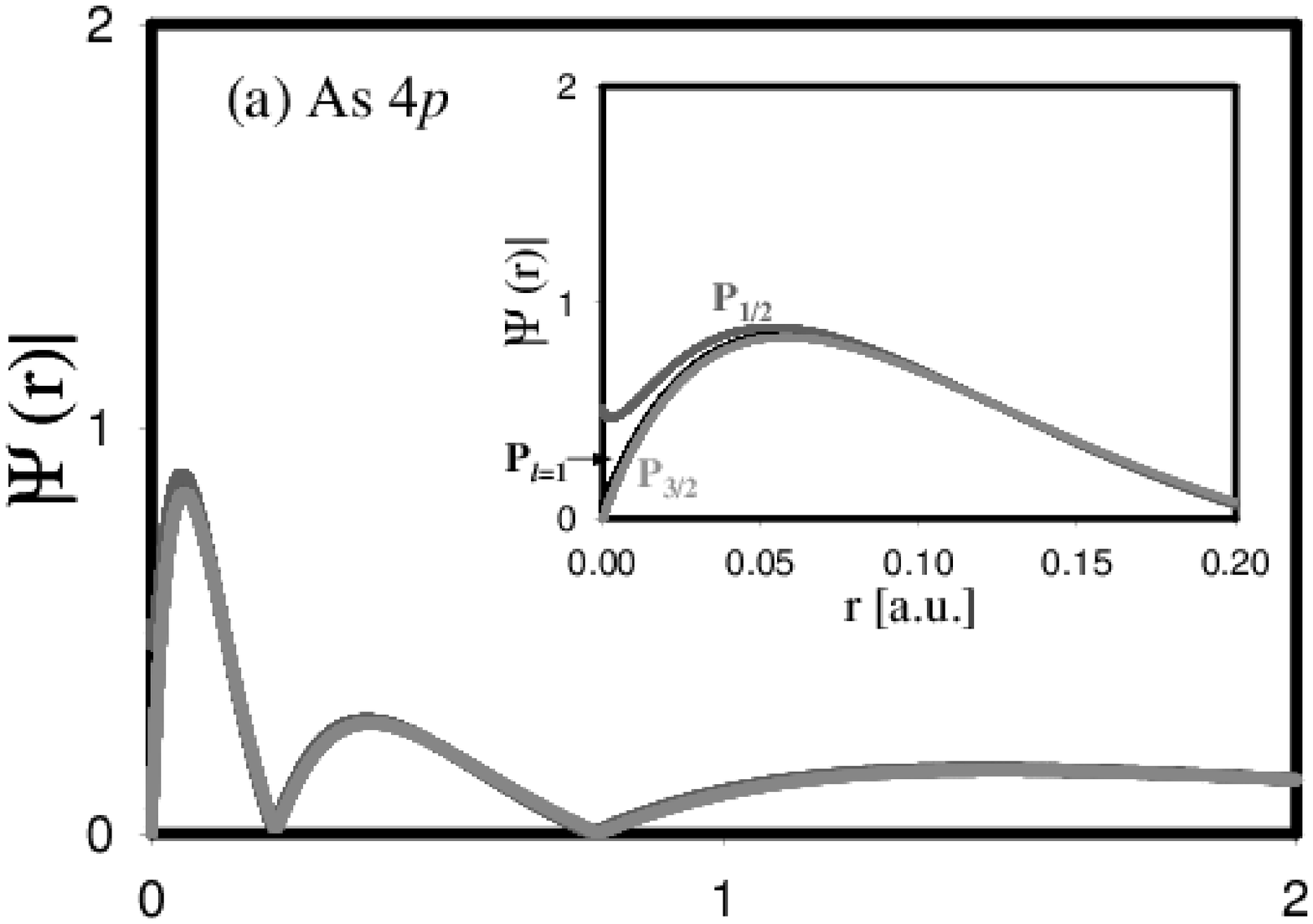}
   \includegraphics*[width=7.5cm]{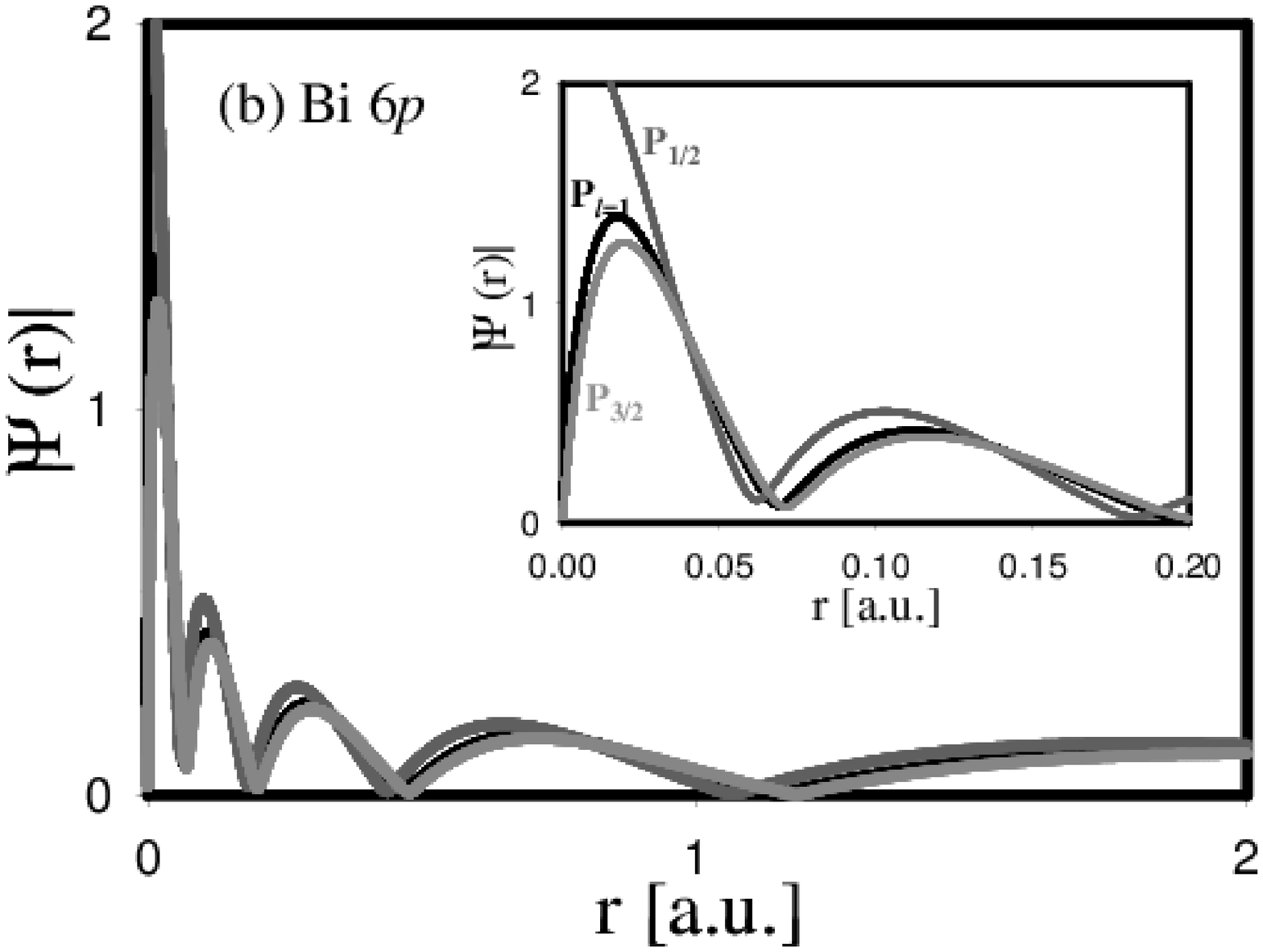}
 \caption
 { 
 Comparison of $p_{1/2}$, $p_{3/2}$, and $p_{l=1}$ orbitals in atomic As and Bi showing the large discrepancy between 
$p_{1/2}$ and 
the $p_{l=1}$ orbitals, especially for the heavier Bi atom.
 }
 \label{AsBi}
 \end{figure}

Spin-orbit (SO) splitting $\Delta_{SO}=\epsilon
(\Gamma_{8v}) - \epsilon (\Gamma_{7v})$ at the top of the valence
band of a semiconductor is an important parameter for the
determination of optical transitions in these systems.\cite{Cardona,Wei94,Wei02} It is also
an important parameter to gauge the chemical environment and bonding
of a semiconductor.\cite{Cardona,Phillips,Wei89,VanVechten,Parayananthal}  Extensive studies of SO splitting, both
theoretically\cite{CardonaSO,Wepfer,CardonaTheo,Poon,Willatzen,AlDouri,Clas,WeiZunger} and 
experimentally,\cite{Madelung,Landolt,Ortner,Parsons,Losch,Jung,Montegu,Hopfield,Kim,Marple,Janowitz,Niles,Herman,Wu,Vurgaftman}
have been carried out in the
past. However, most of these studies focussed on a specific compound
or a small group of similar compounds. Therefore, the general trends of
the spin-orbit splitting in zinc-blende semiconductors is not very
well established. From the experimental point of view, some of the data
were measured more than 30 years ago,\cite{Landolt} and the accuracy of these data
is still under debate. For example, previous
experimental data suggest that CdTe and HgTe have SO splittings $\Delta_{SO}$ at
about 0.8 and 1.08 eV, respectively.\cite{Landolt} These values have been used
widely by experimental groups\cite{Ortner} to interpret optical and magneto-optical
transition data of CdTe, HgTe, and related alloys and heterostructures.
However, recent experimental data suggest that $\Delta_{SO}$ for CdTe
and HgTe are instead around 0.95 eV\cite{Niles} and 0.91 eV.\cite{Janowitz} Without basic
understanding of the general trends of variation of $\Delta_{SO}$ in
tetrahedral semiconductors, it is difficult to judge what should be
the correct value of $\Delta_{SO}$ for CdTe and HgTe.  There are also
several non-conventional II-VI and III-V semiconductors that do not have
a zinc-blende ground state (e.g., CdO, MgO, GaBi, InBi), but that do
form zinc-blende alloys with other compounds, and are currently under
intensive research as novel optoelectronic materials.\cite{Oe,Janotti,Gruber,Tixier}
Therefore, it is important to know the spin-orbit splittings
of these compounds in the zinc-blende phase and understand how they vary as a
function of alloy concentration $x$ in the alloy.
 \begin{figure}[h]
   \includegraphics*[width=7.5cm]{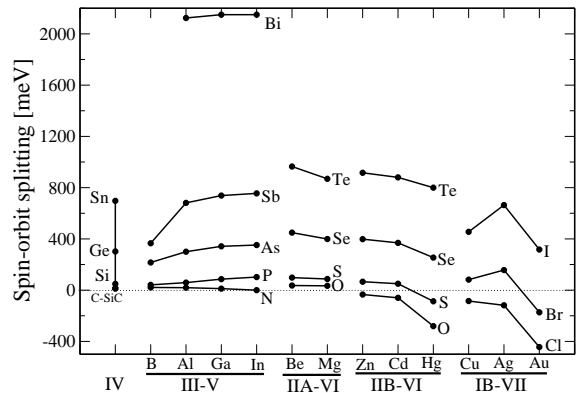}
 \caption
 { 
 Chemical trend of the spin-orbit splittings for all diamond-like group IV and zinc-blende group III-V, II-VI, and I-VII 
semiconductors, including the
 $p_{1/2}$ local orbitals. The graph corresponds to the data  in column ``LAPW+$p_{1/2}$'' of Tables \ref{SOdata435},
 \ref{SOdata26}, and \ref{SOdata17}.
 }
 \label{TrendsFig}
 \end{figure}

From the theoretical point of view, various approximations have been used
to calculate and/or predict SO splitting $\Delta_{SO}$. However, it is not clear
how these approximations affect the calculated $\Delta_{SO}$. For
example, one of the most widely used procedures for calculating the SO
coupling using the density functional theory\cite{Hohenberg} (DFT) and local density
approximation\cite{Kohn,Perdew} (LDA) is the second-variation method\cite{KoellingHarmon,MacDonald} used in many 
all-electron linearized augmented plane wave (LAPW) codes.\cite{WIEN2k,WeiKrakauer,Singh}  In this
approach, following the suggestion of Koelling and Harmon,\cite{KoellingHarmon}
the Hamiltonian of the relativistic Dirac equation is separated into a
``J-weighted-averaged'' scalar relativistic Hamiltonian $H_{SR}$,
in which the dependancy on the quantum number $\kappa$ [where $\kappa=\pm(j + 1/2)$, with $|\vec{j}| = 
|\vec{l} + \vec{s}| =  l \mp 1/2$]  
is removed from the full Hamiltonian,
and a spin-orbit
Hamiltonian $H_{SO}$ with
\begin{eqnarray*}
H_{SO}= \frac{\hbar}{(2Mc)^2}\frac{1}{r}\frac{dV}{dr}(\vec{l}\cdot\vec{s}),\\
\end{eqnarray*}
where 
\begin{eqnarray*}
M = m + \frac{\epsilon-V}{2c^2}
\end{eqnarray*}
is the relativistically enhanced electron mass, $c$ is the speed of
light, V is the effective potential, $\epsilon$ is the eigenvalue,
and $\vec{s}$ and $\vec{l}$ are the Pauli spin and
angular momentum operators, respectively. The scalar relativistic
Hamiltonian, which includes the mass velocity and Darwin corrections, 
is solved first using standard diagonalization method for
each spin orientation (or solved just once if the system is not spin
polarized).  The SO Hamiltonian is included subsequently, in where the
full Hamiltonian is solved using the scalar relativistic wavefunctions
as basis set. Normally, only a small number of scalar relativistic
wavefunctions are included in the second step, and only the spherical
part of the potential within a muffin-tin sphere centered on each atomic
site is used in the SO Hamiltonian. The advantage of the second-variation 
method is the physical transparency (e.g., it keeps spin as
a good quantum number as long as possible) and the efficiency, because,
in most cases in the second step, only a small number of basis
functions are needed to have good agreement with solutions of fully
relativistic Dirac equations. This approach has been shown 
to obtain $\Delta_{SO}$ that is in excellent agreement with experiments. For
example, the calculated $\Delta_{SO}$ for GaAs is 0.34 eV compared
with experimental data of 0.34 eV.\cite{Landolt} However, one major approximation in
the ``J-weighted-averaged'' treatment is the replacement of the two
$p_{1/2}$ and $p_{3/2}$ orbitals by one $p_{l=1}$ orbital. Although
this is a good approximation for atoms with low atomic number, it has
been show that such approximation fails for heavy atoms.\cite{MacDonald,failure,6p12} The main
reason for this failure is because the $p_{1/2}$ orbital has finite
magnitude at the nuclear site, whereas the $l=1$ orbital has zero magnitude at
the nuclear site. 
Figure~\ref{AsBi} plots the $p_{1/2}$, $p_{3/2}$,
and $p_{l=1}$ orbitals for As (Z=33) and Bi (Z=83).
As we can see, the $p_{1/2}$ orbital deviates significantly from the $p_{l=1}$ orbital near the origin.
The error clearly increases as the atomic number increases, and is very large
for heavier elements such as Bi.
Therefore, the $p_{1/2}$ orbital is not very well
represented near the nuclear site using the $p_{l=1}$ orbital, even with
the addition of its energy derivative in the linearization
procedure.\cite{Singh} Consequently, the SO splitting cannot be accurately evaluated, in general,
with solely the $p_{l=1}$ orbital.
However,
no systematic studies have been done to evaluate the effect of the $p_{1/2}$
orbital on the calculated SO splitting $\Delta_{SO}$.
 \begin{figure}[h]
   \includegraphics*[width=7.5cm]{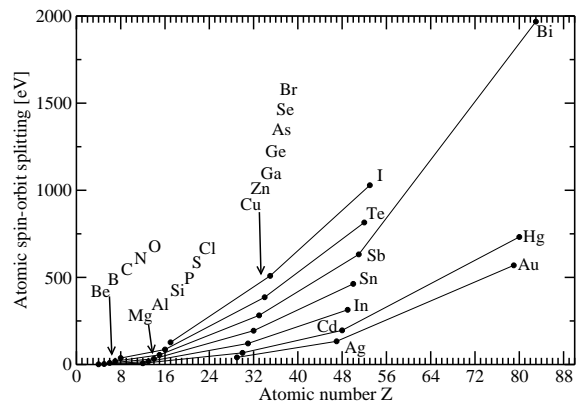}
 \caption
 { 
 Atomic spin-orbit splittings $\epsilon(p_{3/2}) - \epsilon(p_{1/2})$ for atoms studied in this paper.
The spin-orbit splittings increase as a function of 
 the atomic number Z. See Table \ref{p12p32} for data subdivided according to their respective groups.
 }
 \label{Atomic}
 \end{figure}

The objective of this paper is to do a systematic study of the SO
splitting $\Delta_{SO}$ of all diamond-group IV and zinc-blende groups
III-V, II-VI, and I-VII semiconductors using the first-principles band-structure
method within the density functional formalism.  
We find that the calculated SO splittings including the $p_{1/2}$ local orbital 
are in good agreement with available experimental data. The general
chemical trends of the $\Delta_{SO}$ are revealed and explained in terms
of atomic SO splittings, volume effects, and $p$--$d$ coupling effects.

\section{Method of calculations}
The calculations are performed using the full potential linearized
augmented plane wave (FLAPW) method as implemented in the
WIEN2k code.\cite{WIEN2k,Singh}
The frozen core projector augmented wave
(PAW) approach implemented in the VASP code\cite{VASP, Bloechl} is
used for comparison. We used the Monkhost-Pack\cite{Monk} 4 $\times$ 4
$\times$ 4 \textbf{k} points for the Brillouin zone integration. For
the FLAPW method, SO coupling is included using the second-variation
method performed with or without the $p_{1/2}$ local orbitals.  Highly
converged cutoff parameters in terms of the numbers of spherical
harmonics inside the muffin-tin region and the plane waves in the
interstitial region, as well as local orbitals for low-lying valence
band states (anion $s$ and cation $d$ states), are used to ensure the
full convergence of the calculated values.  For the PAW method, high
precision energy cutoffs have been chosen for all semiconductors (as
large as 37 Rydberg for the nitrides and oxides).
\begin{table}[h]
\caption{Calculated spin-orbit splitting $\Delta_{SO}$ for all diamond group IV and
zinc-blende group III-V semiconductors, using the FLAPW method with
or without the $p_{1/2}$ local orbitals and the frozen-core PAW method. Our results are compared with 
available experimental data.
Our error analysis suggests that 
the uncertainty of the LDA calculated value is less than 20 meV.
}
 \begin{tabular}{llcccc} \hline\hline
Comp. & $a$ (\AA)  & \multicolumn{3}{c}{$\Delta_{SO}$ [meV]} \\  \cline{3-5}  \\
& & LAPW &  LAPW+${\textbf p_{\textbf 1/2}}$  & PAW & exper. \\ \hline 
\underline{\bf IV} \\
C                  & 3.5668 &    13 &    13 &    14 &   13$^a$   \\
SiC                & 4.3596 &    14 &    14 &    15 &   10$^b$   \\
Si                 & 5.4307 &    49 &    49 &    50 &  44$^{c}$   \\
Ge                 & 5.6579 &   298 &   302 &   302 & 296$^b$   \\
$\alpha$-Sn        & 6.4890 &   669 &   697 &   689 & 800$^{c}$   \\  \hline
\underline{\bf III-V} \\
BN                 & 3.6157 &    21 &    21 &    22 & ---   \\
BP                 & 4.5383 &    41 &    41 &    42 & ---   \\
BAs                & 4.7770 &   213 &   216 &   212 & ---   \\
BSb                & 5.1982 &   348 &   366 &   346 & ---   \\ \\
AlN                & 4.3600 &    19 &    19 &    19 & 19$^{d}$  \\
AlP                & 5.4635 &    59 &    59 &    62 & ---   \\
AlAs               & 5.6600 &   296 &   300 &   305 & 275$^b$, 300$^c$   \\
AlSb               & 6.1355 &   658 &   681 &   679 & 750$^b$, 673$^c$    \\
AlBi               & 6.3417 & 1 895 & 2 124 & 2 020 & ---   \\ \\
GaN                & 4.5000 &    12 &    12 &    12 &  11$^{c}$, 17$^d$   \\
GaP                & 5.4505 &    86 &    86 &    88 &  80$^{c}$   \\
GaAs               & 5.6526 &   338 &   342 &   342 & 341$^{c}$   \\
GaSb               & 6.0951 &   714 &   738 &   722 & 752$^{c}$, 730$^e$   \\
GaBi               & 6.3240 & 1 928 & 2 150 & 2 070 & ---   \\ \\
InN                & 4.9800 &    -1 &     0 &     0 &   5$^d$   \\
InP                & 5.8687 &   100 &   102 &   104 & 108$^{c}$, 99$^f$   \\
InAs               & 6.0583 &   344 &   352 &   355 & 371$^b$, 380$^c$   \\
InSb               & 6.4794 &   731 &   755 &   754 & 803$^b$, 850$^c$, 750$^g$   \\
InBi               & 6.6860 & 1 917 & 2 150 & 2 089 & ---   \\
\hline\hline
 \end{tabular}
 \label{SOdata435} \\
$^a$ Reference \onlinecite{Serrano}. \hfill \mbox{}  \\
$^b$ Reference \onlinecite{Landolt}. \hfill \mbox{} \\
$^c$ Reference \onlinecite{Madelung}. \hfill \mbox{}  \\
$^d$ Reference \onlinecite{Vurgaftman}. \hfill \mbox{}  \\
$^e$ Reference \onlinecite{Parsons}. \hfill \mbox{} \\
$^f$ Reference \onlinecite{Losch}. \hfill \mbox{} \\
$^g$ Reference \onlinecite{Jung}. \hfill \mbox{}

 \end{table}

In most cases, the band structure calculations are performed at the
experimental lattice constants. For compounds that have only
experimental lattice constant in the wurtzite structure, such as ZnO, we
assume that zinc-blende ZnO has the same volume as in its wurtzite
structure.\cite{Madelung} For BSb, the (Al, Ga, In)Bi, and the (Be, Mg, Cd, Hg)O,
that does not have either zinc-blende or wurtzite experimental
structure parameters, the LDA-calculated lattice constants are used.
For the silver halides and the gold halides, the LDA lattice constants have
been corrected according to the small discrepancy between LDA and experiment values of
AgI (more precisely, 0.088 \AA\ have been added to the LDA lattice constants of
silver halides and gold halides).
The LDA-calculated lattice constants are expected to be reliable. For
example, our predicted\cite{Janotti} lattice constant of GaBi 
is $a=6.324$~\AA, whereas recent experimental observation\cite{Tixier} finds 
a value around 6.33$\pm$0.06 \AA, 
in good agreement with our prediction. All
the lattice constants used in our calculation are listed in Tables \ref{SOdata435}, \ref{SOdata26}, 
and \ref{SOdata17}.
 \begin{figure}[h]
   \includegraphics*[width=7.5cm]{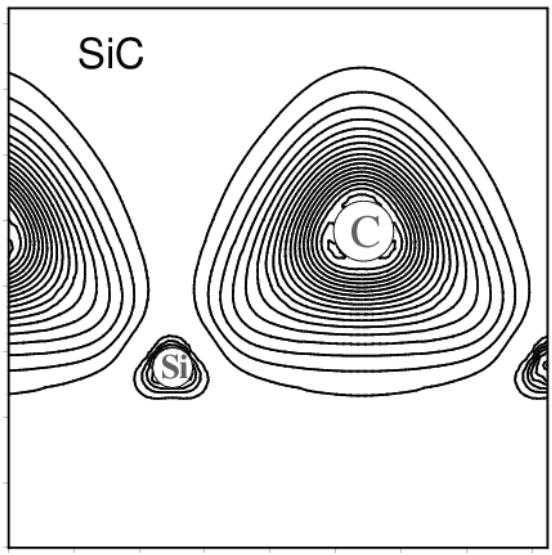}
 \caption{ Charge distribution at the VBM for SiC. The charges are mostly distributed on the carbon atom site.
 }
 \label{SiC}
 \end{figure}

\begin{table}[h]
\caption{Calculated spin-orbit splitting $\Delta_{SO}$ for all 
IIA-VI and IIB-VI semiconductors, using the FLAPW method, with
or without the $p_{1/2}$ local orbitals, and the frozen-core PAW method. 
The lattice constants with a $\star$ corresponds to one at their LDA energy minimum (for ZnO$^{\star\star}$, 
the lattice constant of the zinc-blende
structure is chosen so that its volume is equal to that in the wurtzite structure).
Our results are compared with available experimental data.
Our error analysis suggests that due to the overestimation of the $p$--$d$ hybridization, our calculated
$\Delta_{SO}$ is underestimated by 30, 40, and 110 meV for Zn, Cd, and Hg compounds, respectively.
For other compounds, the LDA error is estimated to be less than 20 meV.
}
 \begin{tabular}{llcccc} \hline\hline
Comp. & $a$ (\AA)  & \multicolumn{3}{c}{$\Delta_{SO}$ [meV]} \\  \cline{3-5}  \\
& & {\tiny LAPW} &  {\tiny  LAPW+${\textbf p_{\textbf 1/2}}$}  & {\tiny PAW} & exper. \\ \hline 
\underline{\bf IIA-VI}  \\
BeO       & 3.7654$^{\star}$      &    36 &    36 &    38 & ---   \\
BeS       & 4.8650                &    98 &    98 &    98 & ---   \\
BeSe      & 5.1390                &   445 &   449 &   447 & ---   \\
BeTe      & 5.6250                &   927 &   965 &   944 & ---   \\ \\
MgO       & 4.5236$^{\star}$      &    34 &    34 &    34 & ---   \\
MgS       & 5.6220                &    87 &    87 &    87 & ---   \\
MgSe      & 5.8900                &   396 &   399 &   396 & ---   \\
MgTe      & 6.4140                &   832 &   869 &   854 &  945$^a$   \\ \hline
\underline{\bf IIB-VI}  \\
ZnO       & 4.5720$^{\star\star}$ &   -34 &   -34 &   -37 &  -4$^{b}$   \\
ZnS       & 5.4102                &    66 &    66 &    64 &  65$^c$, 86$^d$   \\
ZnSe      & 5.6676                &   393 &   398 &   392 & 420$^{c,e}$, 400$^d$  \\
ZnTe      & 6.0890                &   889 &   916 &   898 & 910$^d$, 950$^a$   \\ \\
CdO       & 5.0162$^{\star}$      &   -59 &   -60 &   -58 & ---   \\
CdS       & 5.8180                &    50 &    50 &    46 &  62$^d$, 56$^{b}$  \\
CdSe      & 6.0520                &   364 &   369 &   370 & 416$^d$, 390$^{e}$    \\
CdTe      & 6.4820                &   848 &   880 &   865 & 810$^c$, 800$^d$, 900$^f$  \\ \\
HgO       & 5.1566$^{\star}$      &  -285 &  -281 &  -292 & ---   \\
HgS       & 5.8500                &  -100 &   -87 &  -108 & ---   \\
HgSe      & 6.0850                &   235 &   254 &   238 & 450$^c$, 396$^d$ 300$^g$  \\
HgTe      & 6.4603                &   762 &   800 &   781 & 1080$^c$, 910$^g$   \\
\hline\hline
 \end{tabular}
 \label{SOdata26} \\
$^a$ Reference \onlinecite{Montegu}. \hfill \mbox{} \\
$^b$ Reference \onlinecite{Hopfield}. \hfill \mbox{} \\
$^c$ Reference \onlinecite{Landolt}. \hfill \mbox{} \\
$^d$ Reference \onlinecite{Madelung}. \hfill \mbox{} \\
$^e$ Reference \onlinecite{Kim}.\hfill \mbox{} \\
$^f$ Reference \onlinecite{Marple}. \hfill \mbox{}  \\
$^g$ Reference \onlinecite{Janowitz}. \hfill \mbox{}
 \end{table}

 \begin{figure}[h]
   \includegraphics*[width=7.5cm]{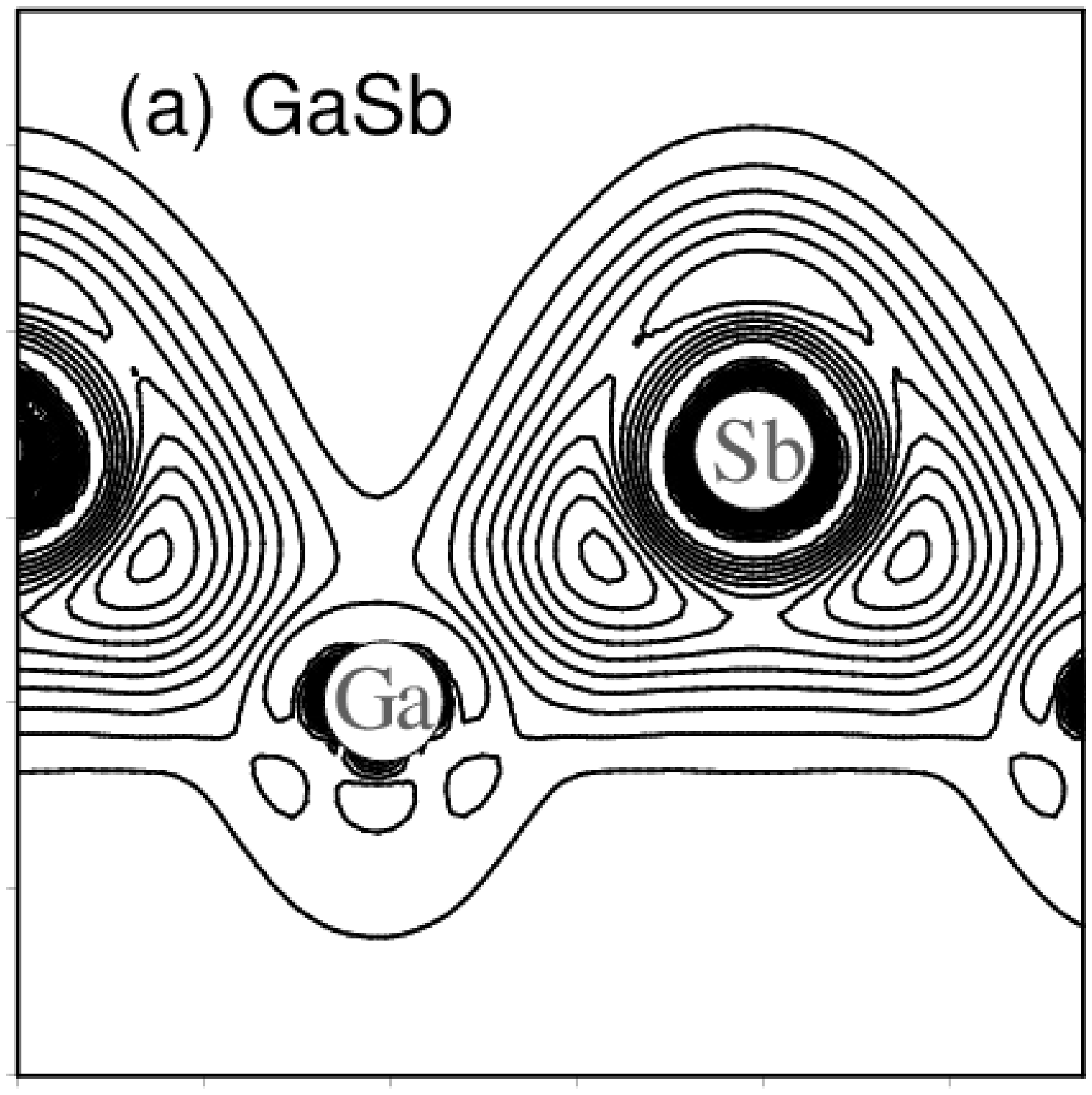}
   \includegraphics*[width=7.5cm]{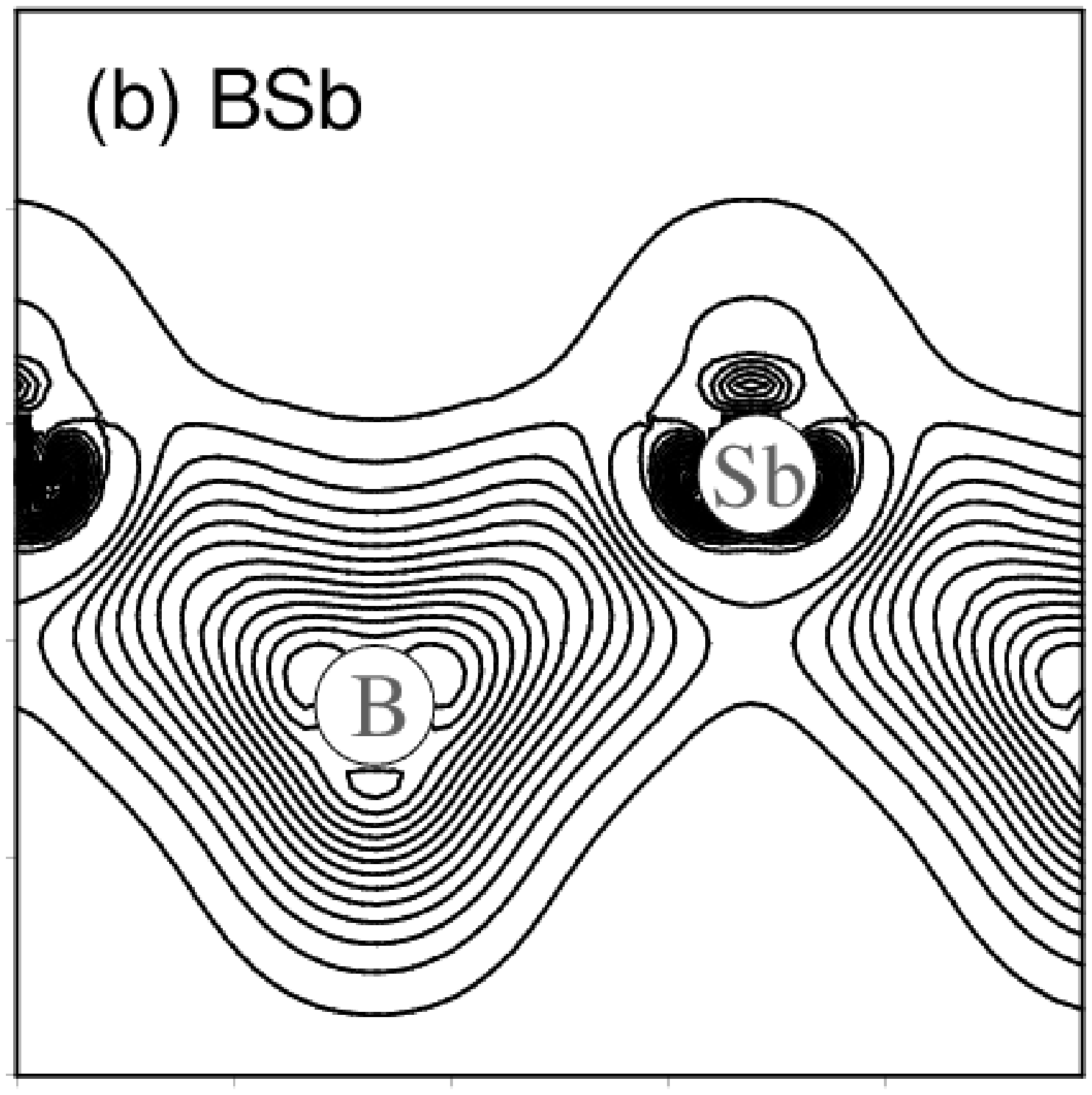}
 \caption
 { 
Charge density of the VBM state for GaSb and BSb, showing that for BSb the role of cation and anion is reversed.
 }
 \label{TheSb}
 \end{figure}

\section{Effect of the $p_{1/2}$ local orbital}
Tables \ref{SOdata435}, \ref{SOdata26}, and \ref{SOdata17} present the calculated SO splittings data for
all diamond-like group IV and zinc-blende groups III-V, II-VI, and I-VII
semiconductors.  The calculated values are obtained with or without the
$p_{1/2}$ local orbitals. We find that including the
$p_{1/2}$ local orbital provides a better variation basis for the
$\Gamma_{7v}$ state, lowers the eigen energy, and, therefore, increases the
SO splitting $\Delta_{SO}$ = $\epsilon$($\Gamma_{8v}$) - $\epsilon$($\Gamma_{7v}$).
The correction due to the $p_{1/2}$ orbital increases as the atomic
number increases. Since the VBM consists of mostly anion $p$ state, the
dependence is more on anion atomic numbers.  
We find that corrections
due to the inclusion of the $p_{1/2}$ local orbital (for both anions and cations) 
are negligible for
lighter atoms, are $\sim$10 meV for 4$p$ anions, $\sim$40 meV for 5$p$
anions and can be as large as $\sim$250 meV for 6$p$ anions. Thus, for Bi
compounds (AlBi, GaBi, and InBi), large errors could be introduced if the
$p_{1/2}$ local orbital is not included.\cite{Janotti}  In all
the cases, inclusion of $p_{1/2}$ local orbital brings a better agreement
between the calculated $\Delta_{SO}$ and available experimental data.
\begin{table}[h]
\caption{Calculated spin-orbit splitting $\Delta_{SO}$ for all 
IB-VII compounds, using the FLAPW method, with
or without the $p_{1/2}$ local orbitals, and the frozen-core PAW method. Our results are compared with available experimental data.
We use experimental lattice constants\cite{Blacha,Ma,Hull} for 
CuX (X= Cl, Br, I) and AgI.
The lattice constants for the other AgX and AuX compounds are estimated from calculated LDA lattice constans and experimental lattice constant
of AgI.
Due to the overestimation of the $d$ character in the VBM, the LDA underestimate the $\Delta_{SO}$ by 20, 60, and 170 meV
for chlorides, bromides, and iodides, respectively.
}
 \begin{tabular}{llcccc} \hline\hline
Compound & $a$ (\AA)  & \multicolumn{3}{c}{$\Delta_{SO}$ [meV]} \\  \cline{3-5}  \\
& & LAPW &  LAPW+${\textbf p_{\textbf 1/2}}$  & PAW & exper. \\ \hline 
\underline{\bf IB-VII}  \\
CuCl     & 5.4057           &   -85 &   -85 &   -85 &  -69$^a$ \\
CuBr     & 5.6905           &    80 &    82 &    86 &  147$^a$ \\
CuI      & 6.0427           &   440 &   455 &   466 &  633$^a$ \\  \\
AgCl     & 5.8893$^{\star}$ &  -119 &  -118 &  -122 &  ---   \\
AgBr     & 6.1520$^{\star}$ &   155 &   157 &   158 &  ---   \\
AgI      & 6.4730           &   643 &   664 &   658 &  837$^a$  \\  \\
AuCl     & 5.7921$^{\star}$ &  -444 &  -444 &  -446 &  ---  \\
AuBr     & 6.0517$^{\star}$ &  -177 &  -173 &  -178 &  ---   \\
AuI      & 6.3427$^{\star}$ &   294 &   317 &   317 &  ---  \\ \hline\hline
 \end{tabular}
 \label{SOdata17} \\
$^a$ References \onlinecite{Cardona63} and \onlinecite{BlachaSSC}. \hfill \mbox{} \\
 \end{table}

\section{Chemical trends}
Figure \ref{TrendsFig} shows the general chemical trends of the
calculated SO splittings $\Delta_{SO}$ for all diamond-like group IV
and zinc-blende III-V, II-VI, and I-VII semiconductors, with inclusion of the $p_{1/2}$ local orbitals.
We find that (i) the
SO splittings increase monotonically when anion atomic number
increases; (ii) the SO splittings increase with the cation atomic
number when the compound is more covalent, such as in
most III-V compounds; (iii) the SO splittings decrease with the cation
atomic number when the compound is more ionic, such as in II-VI and the
III-nitride compounds; (iv) for compounds with the same principal quantum 
number, $\Delta_{SO}$ increases as the ionicity of the compounds increases.
Finally, (v) the halides (IB-VII) constitute a special case because the VBM in IB-VII is no longer an
anion $p$ dominant state.\cite{Blacha}
Therefore, IB-VII compounds do not follow the rules discussed above.

To understand these chemical trends, we will first discuss the factors
that can affect the SO splitting $\Delta_{SO}$ for the systems studied here. 
\textbf{(a) Dependence on the atomic number.} The atomic SO splitting between the $p_{3/2}$ and
$p_{1/2}$ states increases as a function of atomic number $Z$.
Table \ref{p12p32} gives the calculated splitting of the atomic fine structures,
$\epsilon$($p_{3/2}$) - $\epsilon$($p_{1/2}$), as a function of the atomic number Z in their respective groups.
Figure \ref{Atomic} (related to Table \ref{p12p32}) shows the variation of the 
atomic spin-orbit splittings as a function of the atomic numbers,
for all atoms considered.
The spin-orbit splittings increase with the atomic number, as expected.\cite{Baym}
The increases follow approximately a power law with
$\Delta_{SO}(p_{3/2}-p_{1/2}) \propto Z^{\alpha}$, where $\alpha$ is close to 2.
\textbf{(b) Dependence on the volume.} As the volume of the compound
decreases, the charge distribution in the crystal is
renormalized. The bonds become more covalent. More charge is pushed
into a region near the nuclei.
Because the SO coupling is
larger near the nuclear site,
the SO splitting $\Delta_{SO}$ usually
increases as the volume decreases. \textbf{(c) Dependence on the cation valence $d$
orbital.}  The VBM in a majority of zinc-blende semiconductors consists of mostly
anion $p$ and a smaller amount of cation $p$ orbitals. By symmetry, the
VBM state in zinc-blende structure can couple with the cation
$t_{2d}$ orbitals. The cation $t_{2d}$ orbital has a negative contribution\cite{Cardona, WeiZunger}
to the SO splitting $\Delta_{SO}$
(i.e., the $\Gamma_{8v}$ is below the $\Gamma_{7v}$ state).
Thus, large mixing of heavy cation $d$
orbitals in VBM can reduce $\Delta_{SO}$.
 \begin{figure}[htb]
   \includegraphics*[width=4cm]{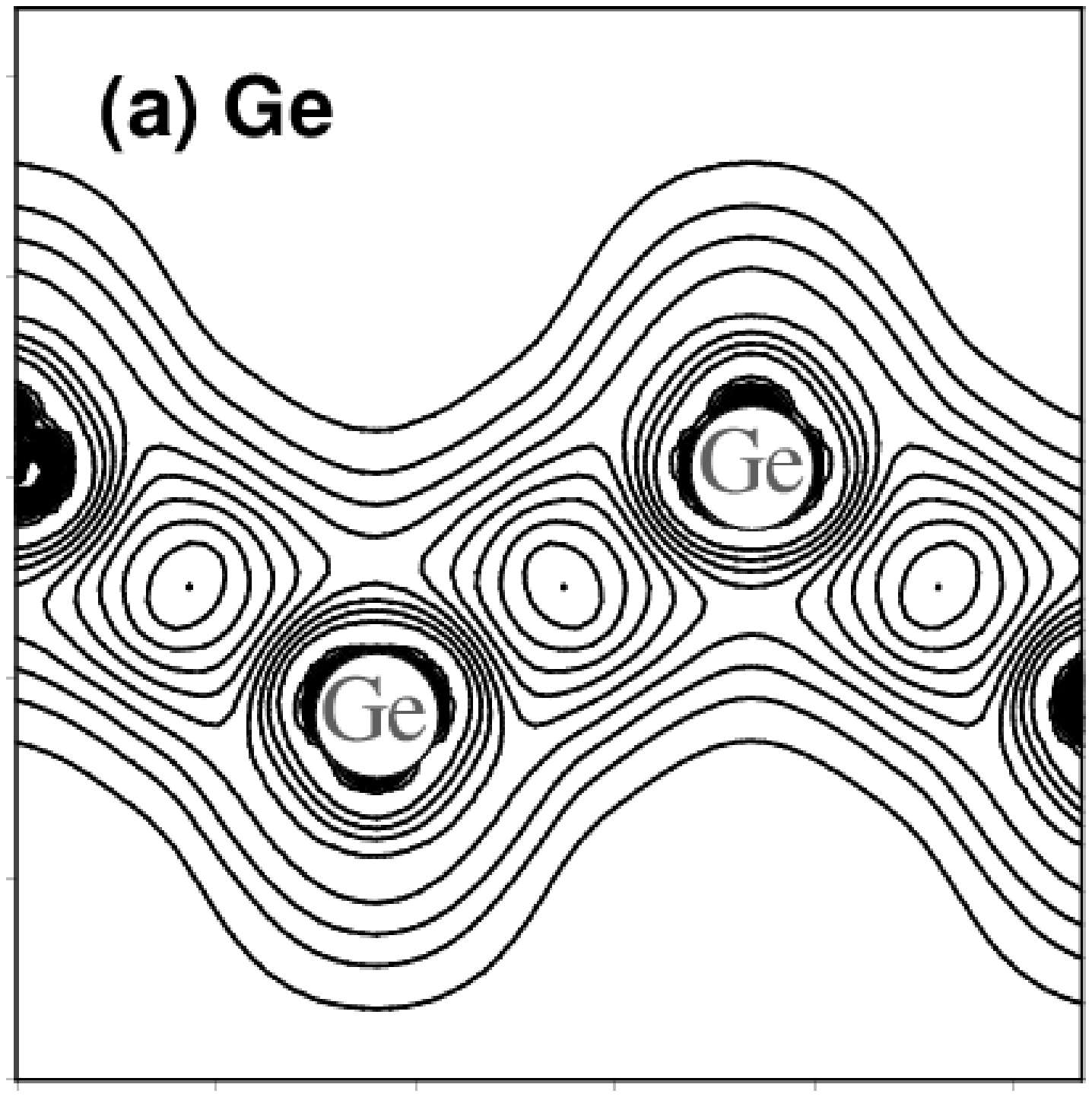}
   \includegraphics*[width=4cm]{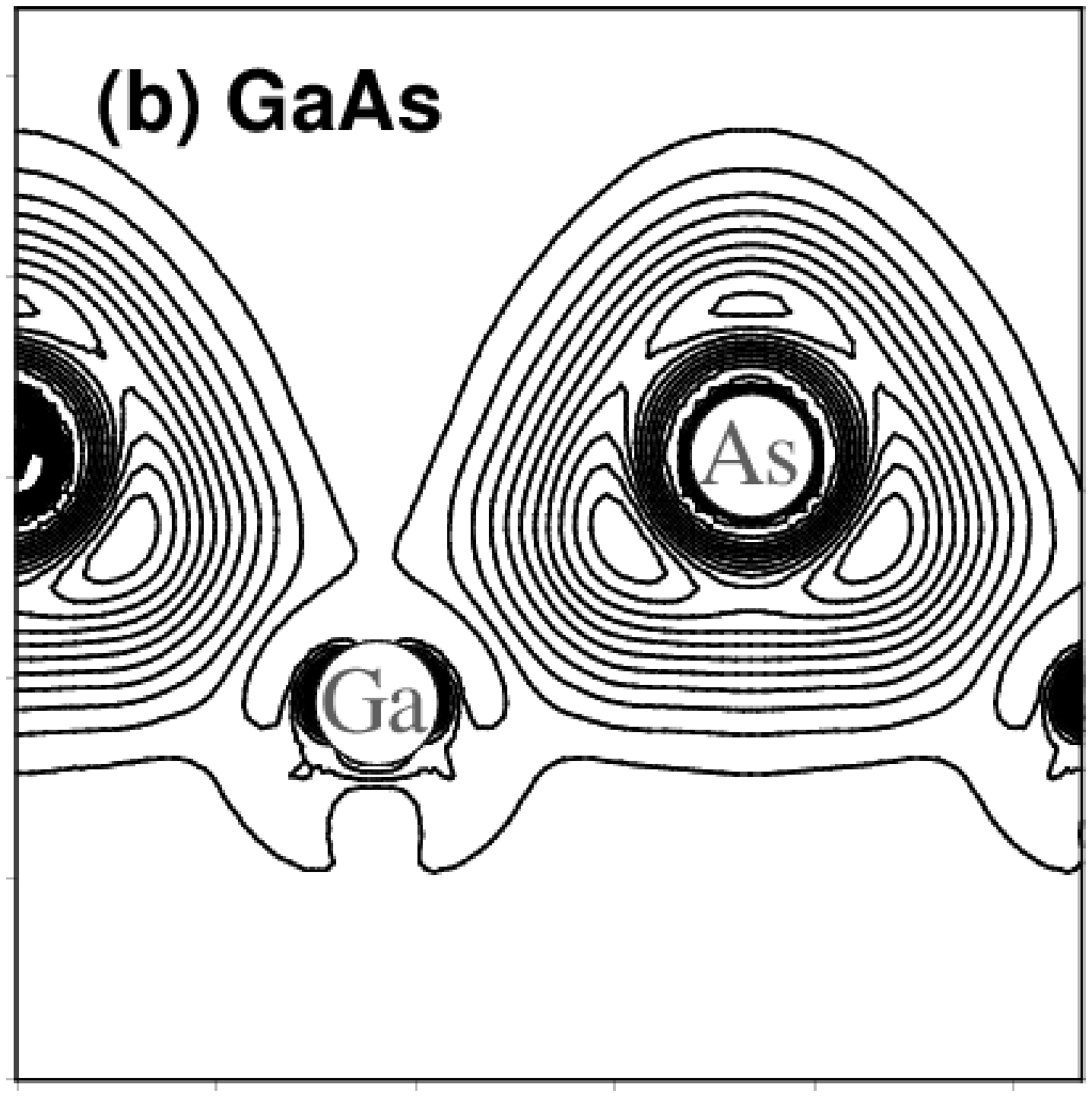}
   \includegraphics*[width=4cm]{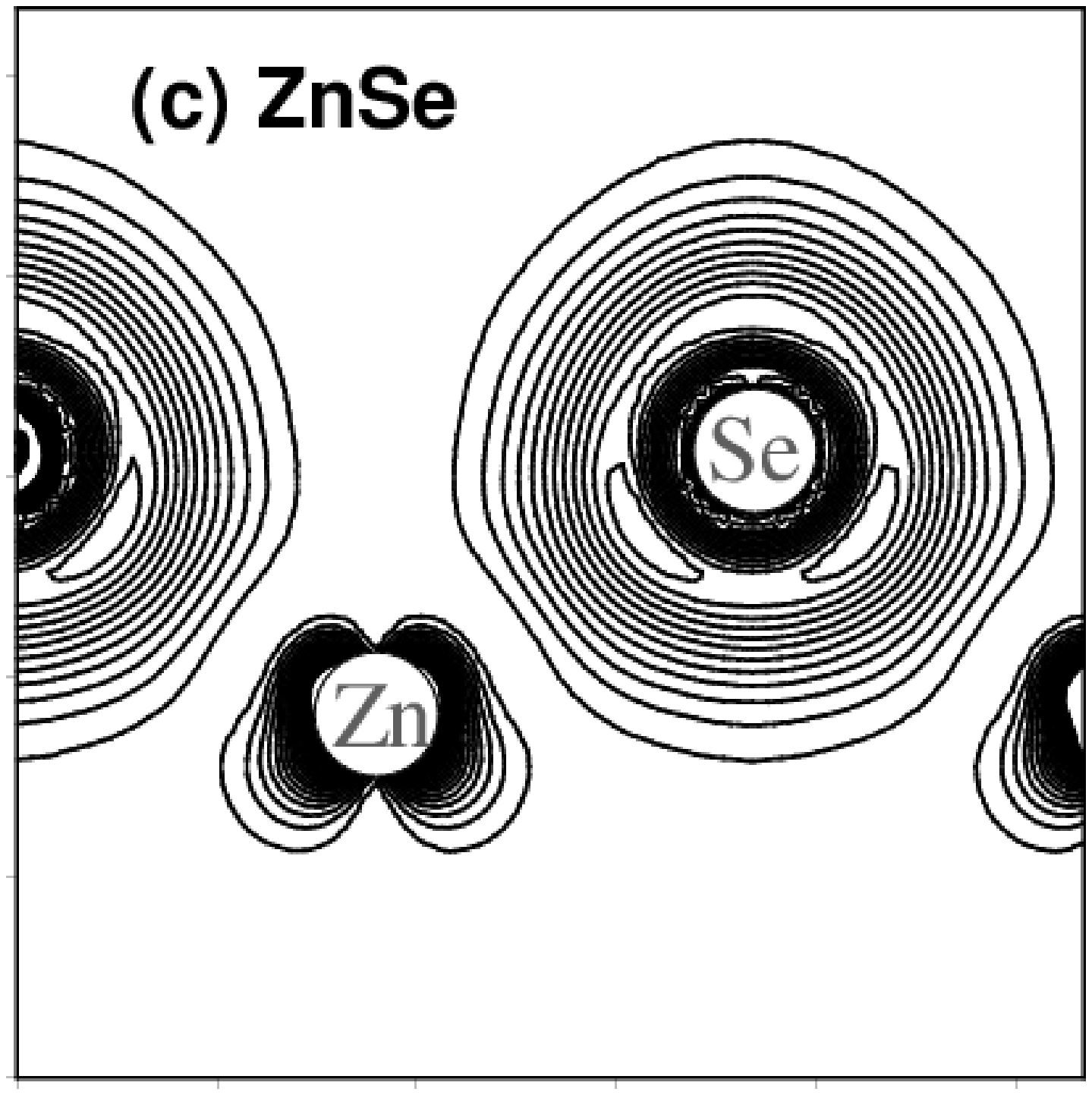} 
   \includegraphics*[width=4cm]{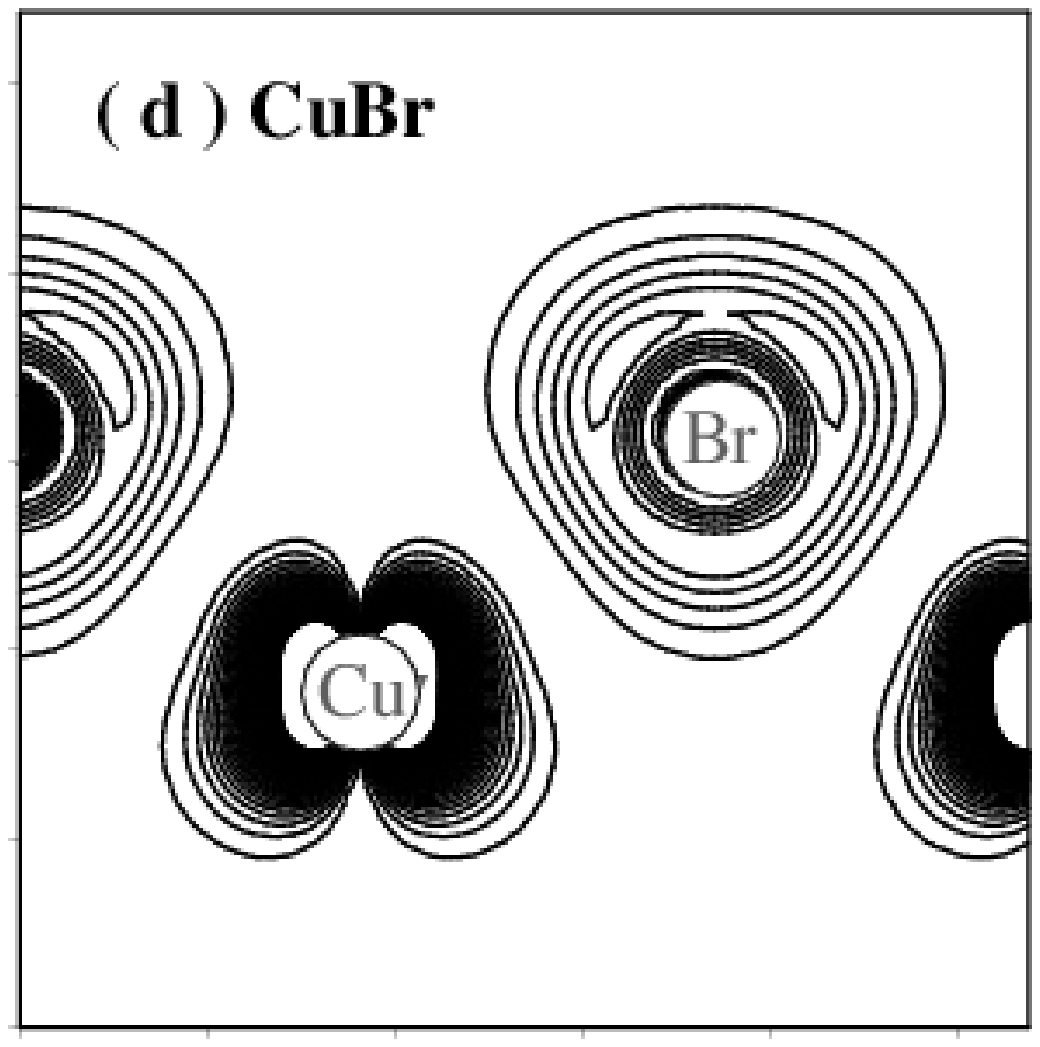}
 \caption
 { 
Charge density of the VBM states for Ge, GaAs, ZnSe, and CuBr showing that as ionicity increases, the charge is more 
localized on the anion site. For ZnSe and CuBr, it also shows antibonding $d$ character on the Zn and Cu sites,
respectively.\cite{WeiZunger}
 }
 \label{GeGaAsZnSe}
 \end{figure}

Using the discussion above, we can now understand the general chemical
trends of the SO splitting $\Delta_{SO}$.

(i) The SO splittings increase monotonically when anion atomic number
increases. For example, $\Delta_{SO}$ increases from 13 $\rightarrow$
49 $\rightarrow$ 302 $\rightarrow$ 697 meV when the atomic number
increases from C $\rightarrow$ Si $\rightarrow$ Ge $\rightarrow$ $\alpha$-Sn;
from 12 $\rightarrow$ 86 $\rightarrow$ 342 $\rightarrow$ 738
$\rightarrow$ 2150 meV when the anion atomic number increases from GaN
$\rightarrow$ GaP $\rightarrow$ GaAs $\rightarrow$ GaSb $\rightarrow$
GaBi; from $-60$ $\rightarrow$ 50 $\rightarrow$ 369 $\rightarrow$
880 meV when the anion atomic number increases from CdO $\rightarrow$ CdS
$\rightarrow$ CdSe $\rightarrow$ CdTe;
from -85 $\rightarrow$ 82 $\rightarrow$ 455 when the anion atomic number increases from
CuCl $\rightarrow$ CuBr $\rightarrow$ CuI.  This is because the VBM has
large anion $p$ character, and the atomic SO splitting of the anion valence
$p$ state increases with the atomic number (see Table \ref{p12p32}). 
One of the interesting case is
SiC. The calculated $\Delta_{SO}$ of 14 meV for SiC is very close to the one of
diamond (13 meV), indicating that SiC is a very ionic material with its VBM containing
mostly C character. Figure \ref{SiC} depicts the contour plot of the charge
distribution at the VBM
for SiC, which shows that the VBM charge is located on the carbon atom site.

(ii) The SO splittings increase with the cation atomic number when the
compound is more covalent, such as in most III-V
compounds. For example, $\Delta_{SO}$ increases from 216 $\rightarrow$
300 $\rightarrow$ 342 $\rightarrow$ 352 meV when the atomic number
increases from BAs $\rightarrow$ AlAs$\rightarrow$ GaAs $\rightarrow$
InAs; from 366 $\rightarrow$ 681 $\rightarrow$ 738
$\rightarrow$ 755 meV when the atomic number increases from BSb
$\rightarrow$ AlSb $\rightarrow$ GaSb $\rightarrow$ InSb. 
This is
because for covalent III-V compounds, the VBM contains significant
amount of cation $p$ orbitals. Therefore, when the cation atomic
number increases, the SO splitting $\Delta_{SO}$ also increases. It is
interesting to note that $\Delta_{SO}$ for BX (X=P, As, and Sb) is
significantly smaller than that for their corresponding common-anion
compounds. 
For example, $\Delta_{SO}$(BSb)=366 meV is only about half of the value of
$\Delta_{SO}$(GaSb)=738 meV. 
This is because boron is much more electronegative than other group III
elements. 
Thus, BX compounds are much more covalent than the other III-V semiconductors.
Figure \ref{TheSb} compares
the charge distribution of the VBM states for BSb and GaSb.
We see that for GaSb, most of the VBM charge is on Sb atom site, whereas for BSb, a large portion of 
the VBM charge is on the B atom site.
Because boron has a small atomic number (Z=5), the SO splitting of B
$2p$ states is very small, leading to very small $\Delta_{SO}$ for
BX. 
This indicates that the
common-anion rule, which states that the variation of $\Delta_{SO}$ is
small for common anion systems, does not apply to all BX, which are extremely
covalent.
 \begin{table}[h]
 \caption{Atomic SO splitting $\epsilon$($p_{3/2}$) - $\epsilon$($p_{1/2}$) for the compounds of Tables~\ref{SOdata435}, 
\ref{SOdata26}, and \ref{SOdata17}, according to their atomic groups.
The data are also depicted in Fig.~\ref{Atomic}, as a function of atomic numbers Z. }
 \begin{tabular}{lcccc} \hline\hline
element      & 
\mbox{} \hspace{2mm} \mbox{}& atomic number & \mbox{} \hspace{2mm} \mbox{}
&    $\epsilon$($p_{3/2}$) - $\epsilon$($p_{1/2}$) [meV]\\ \hline
\underline{\bf IB}   \\
Cu   & & Z=29     & &       41    \\
Ag   & & Z=47     & &      133    \\
Au   & & Z=79     & &      569    \\ \hline
\underline{\bf IIA}   \\
Be   & & Z=4      & &        1    \\
Mg   & & Z=12     & &        7    \\  \hline
\underline{\bf IIB}   \\
Zn   & & Z=30     & &       67    \\
Cd   & & Z=48     & &      196    \\
Hg   & & Z=80     & &      732    \\ \hline
\underline{\bf III}   \\
B   & &  Z= 5     & &        3    \\
Al   & & Z=13     & &       17   \\
Ga   & & Z=31     & &      121   \\
In   & & Z=49     & &      314    \\  \hline
\underline{\bf IV}      \\
C  & &   Z= 6     & &        9     \\
Si  & &  Z=14     & &       33    \\
Ge  & &  Z=32     & &      194   \\
Sn   & & Z=50     & &      463   \\  \hline
\underline{\bf V}   \\
N    & & Z= 7     & &       19     \\
P    & & Z=15     & &       55    \\
As   & & Z=33     & &      282   \\
Sb   & & Z=51     & &      632    \\
Bi   & & Z=83     & &    1 968 \\ \hline
\underline{\bf VI}   \\
O    & & Z= 8     & &       37    \\
S    & & Z=16     & &       86    \\
Se   & & Z=34     & &      386   \\
Te   & & Z=52     & &      815    \\  \hline
\underline{\bf VII}   \\
Cl   & & Z=17     & &      127    \\
Br   & & Z=35     & &      509    \\
I    & & Z=53     & &    1 029    \\  \hline\hline
 \end{tabular}
 \label{p12p32}
 \end{table}

(iii) The SO splittings decrease with the cation atomic number when
the compound is more ionic, such as in II-VI and III-nitride
compounds. For example, $\Delta_{SO}$ decreases from 449 $\rightarrow$
399 meV when the atomic number increases from BeSe $\rightarrow$ MgSe;
from 965 $\rightarrow$ 869 meV when the atomic number increases from
BeTe $\rightarrow$ MgTe; from 398 $\rightarrow$ 369 $\rightarrow$ 254
meV when the atomic number increases from ZnSe $\rightarrow$ CdSe
$\rightarrow$ HgSe; from 21 $\rightarrow$ 19 $\rightarrow$ 12
$\rightarrow$ 0 meV when the atomic number increases from BN to AlN
$\rightarrow$ GaN $\rightarrow$ InN.  This is because for ionic II-VI 
and III-nitride systems, the VBM is mostly an anion $p$ state, thus the
$\Delta_{SO}$ is not sensitive to the cation atomic number or potential.  
However, when cation atomic number decreases, say from Mg
to Be, the volume of the compounds decreases (Table \ref{SOdata26}), and therefore, due
to the charge renormalization effect, the $\Delta_{SO}$ increases. In particular, for the IIB-VI
systems and the III-nitrides, the coupling between cation $d$ and anion
$p$ also plays an important role in the observed trend, because the
$p$--$d$ hybridization is significant in these systems (See Fig.~\ref{GeGaAsZnSe}c). The $p$--$d$
hybridization reduces $\Delta_{SO}$,\cite{Cardona,WeiZunger} and the effect increases when
cation atomic number increases. This explains why $\Delta_{SO}$(HgX)
(for X=O, S, Se, Te) is smaller than $\Delta_{SO}$(CdX), even though they
have similar volume, and why $\Delta_{SO}$(InN) is smaller than
$\Delta_{SO}$(GaN). Note that negative $\Delta_{SO}$ can exist in some
of the compounds such as ZnO, CdO, and HgO where the anion is
light, so their $p$ orbitals have only a small contribution to
$\Delta_{SO}$, but the negative contribution of the cation $d$ orbital is
large.

(iv) For compounds with the same principal quantum number $n$, $\Delta_{SO}$
increases as the ionicity of the compound increases. For example, for
$n=2$, from C $\rightarrow$ BN $\rightarrow$ BeO, the SO splittings
$\Delta_{SO}$ increase from 13 $\rightarrow$ 21 $\rightarrow$ 36 meV;
for $n=3$, from Si $\rightarrow$ AlP $\rightarrow$ MgS, the SO
splittings increase from 49 $\rightarrow$ 59 $\rightarrow$ 87 meV; for
$n=4$, from Ge $\rightarrow$ GaAs $\rightarrow$ ZnSe, the SO
splittings increase from 302 $\rightarrow$ 342 $\rightarrow$ to 398
meV; for $n=5$, from $\alpha$-Sn $\rightarrow$ InSb $\rightarrow$ CdTe, the SO
splittings increase from 697 $\rightarrow$ 755 $\rightarrow$ 880 meV.
The reason for this increase can be understood from plots in Fig.~\ref{GeGaAsZnSe}, which 
show the charge distribution of the VBM states of Ge, GaAs, and ZnSe.
As the system changes from
group IV $\rightarrow$ III-V $\rightarrow$ II-VI, the compound
becomes more ionic and the VBM becomes more localized on the anion
site with increasing atomic number, thus $\Delta_{SO}$ increases.
It is interesting to note that the differences of $\Delta_{SO}$
between the II-VI, the III-V, and the group IV compounds in the same row
increases as $n$ increases (almost doubles when $n$ increases by
one). This is explained by the fact that the atomic number $Z$ almost doubles when $n$ is
increased by one, whereas the atomic SO splitting is proportional to $Z^\alpha$
with $\alpha$ close to 2 (See Table \ref{p12p32} and the discussion above); thus, 
the difference is proportional to $Z$.

(v) The A$^{\mbox{\scriptsize IB}}$X$^{\mbox{\scriptsize VII}}$ halides 
(A$^{\mbox{\scriptsize IB}}$ = Cu, Ag, Au; X$^{\mbox{\scriptsize VII}}$ = Cl, Br, I)
constitute a group of special compounds that do not follow the rules discussed above. 
For example, when moving from ZnSe to CuBr with increased ionicity (see Figure~\ref{GeGaAsZnSe}), 
the SO splitting of CuBr (82 meV) is much smaller than that for
ZnSe (398 meV).
The SO splitting of AgI (664 meV) is also much smaller than that of CdTe (880 meV).
Furthermore, many of the IB-VII compounds (CuCl, AgCl, AuCl, and CuBr) have negative SO splittings, and for these ionic compounds
CuX$^{\mbox{\scriptsize VII}}$ has much smaller SO splittings than AgX$^{\mbox{\scriptsize VII}}$ and AuX$^{\mbox{\scriptsize VII}}$.
The origin of these anomalies is due to the fact that for most of the IB-VII compounds the VBM is no longer an anion $p$ dominated state.
Instead, they are cation $d$ states strongly hybridized with the anion $p$ state.
For instance, in Figure~\ref{GeGaAsZnSe}(d) we show that the VBM of CuBr has a very pronounced antibonding $d$ character at the cation Cu site.
Because the $d$ state has negative $\Delta_{SO}$, this explains why some of the IB-VII compounds have negative $\Delta_{SO}$.
Furthermore, because Cu 3$d$ level is much higher than Ag 4$d$ and Au 5$d$ levels, the VBM of Cu halides contains more cation $d$
character than Ag and Au compounds.
This explains why Cu halides have much smaller $\Delta_{SO}$ than the Ag and Au common anion halides.

\section{Comparison with experiments}

Our calculated results with the $p_{1/2}$ local orbitals are compared with
experimental 
data.\cite{Madelung,Landolt,Ortner,Parsons,Losch,Jung,Montegu,Hopfield,Kim,Marple,Janowitz,Niles,Herman,Wu,Vurgaftman}
For most semiconductors the agreement is very good.
For example, the calculated value for diamond (13 meV) is in very good agreement with the
recent experimentally derived value of 13 meV.\cite{Serrano} 
The experimental value for SiC
in the zinc-blende structure (10 meV)\cite{Landolt,Clas} is smaller than the one for C, therefore, does not follow the chemical trend.
We suggest that the measured value is possibly underestimated.
For most semiconductors,
the difference between theory and experiment is usually less than 20 meV. However, there are several
noticeable cases in which the difference is much larger. For example, for $\alpha$-Sn,
the calculated value is 697 meV, whereas the value in experiment data\cite{Madelung} is
$\sim$800 meV.  
For HgTe the calculated value at
800 meV is much smaller than the widely used 
experimental value\cite{Landolt} of 1080 meV. 
To understand the origin of the discrepancy, we performed the following 
tests. First, we considered a different numerical approach,
i.e., the frozen core PAW method as implemented in
the VASP code to calculate the SO splitting $\Delta_{SO}$.  
Despite the large
difference in the way the SO coupling is implemented in the calculations,
we find that the $\Delta_{SO}$ calculated with the PAW method are very similar to that obtained
with the FLAPW method.  For $\alpha$-Sn and HgTe, the results obtained by
the PAW method are 689 and 781 meV, respectively, in good agreement with
the FLAPW-calculated values of 697 and 800 meV. Next, we estimated
the effect of $p$--$d$ coupling. It has been argued that the
LDA-calculated cation $d$ orbitals are too shallow,\cite{WeiZunger} so 
$p$--$d$ hybridization at the VBM is
overestimated, which may lead to smaller calculated $\Delta_{SO}$. To
verify if this is the possible reason, we performed the following
calculations. (i) After obtaining the converged LDA potential, we
removed the cation $d$ orbital from the basis set to calculate the
$\Delta_{SO}$. We find that for $\alpha$-Sn, this procedure has no effect on the
calculated $\Delta_{SO}$. This is consistent with the fact that for
this compound, the cation $d$ and anion $p$ separation is large
enough that the amount of cation $d$ orbital at the VBM is not sufficient to
affect the calculated $\Delta_{SO}$. For ZnTe, CdTe, and HgTe,
removing the cation $d$ orbital increases the $\Delta_{SO}$ by 48, 63,
and 253 meV, respectively. These values are the upper limit on the possible
effect of $p$--$d$ coupling on the calculated $\Delta_{SO}$. (ii) To
get more reliable estimates on the LDA error of the calculated
$\Delta_{SO}$, we added an external potential\cite{WeiCarrier} on the cation
muffin-tin sphere to push down the cation $d$ orbitals such that the calculated
cation binding energy is close to the experimental photemission
data.\cite{WeiZunger}
In this case, the calculated $\Delta_{SO}$ is 0.94, 0.91, and
0.90 eV for ZnTe, CdTe, and HgTe, respectively. The above
analysis demonstrates that the possible LDA error in calculating
$\Delta_{SO}$ is less than 30, 40, and 110 meV for Zn, Cd, and Hg
compounds, respectively, and much smaller for other compounds.

Our analysis above suggests that $\Delta_{SO}$ for $\alpha$-Sn and HgTe
should be around 0.70 and 0.90 eV, respectively, smaller than the
experimental values of 0.80 and 1.08 eV, respectively. The
origin of this discrepancy is still not very clear. But we notice that 
$\alpha$-Sn and HgTe are semimetals,
i.e., the $\Gamma_{6c}$ state is below the VBM.  This makes the
accurate measurement of the $\Delta_{SO}$ for these compounds
more challenging.  Indeed, recent measurements\cite{Janowitz} of $\Delta_{SO}$ for HgTe show
that it has a value of 0.9 eV, in good agreement with our
predicted value. 
We also notice that the recent reported experimental SO splitting for InSb,\cite{Jung} which has a very small band gap 
(0.24 eV), agrees well with our calculation.
Further experimental studies are needed to clarify
these issues.

\section{Summary}
In summary, we have studied systematically the SO splitting $\Delta_{SO}$ 
of all diamond-like group IV  and zinc-blende group III-V, II-VI, and I-VII semiconductors 
using the first-principles band structure method. We studied the 
effect of the $p_{1/2}$ local orbitals on the calculated $\Delta_{SO}$. 
The general trends of $\Delta_{SO}$ of the semiconductors are revealed
and explained in terms of atomic SO splitting, volume deformation-induced 
charge renormalization, and cation-anion $p$--$d$ couplings. 
In most cases, our calculated results are in good agreement with the experimental data.
The differences between our calculated value for $\alpha$-Sn and HgTe, and to a lesser
degree for InAs and GaSb, are highlighted. Experiments are called for to test our predictions.

\section{Acknowledgments}
This work was supported by the U.S. Department of Energy, Grant No. DE-AC36-99G010337.

\end{document}